\documentclass[fleqn,usenatbib]{mnras}
\usepackage{newtxtext,newtxmath}

\usepackage{amsmath}	
\usepackage[T1]{fontenc}
\usepackage{ae,aecompl}
\usepackage{graphicx}	
\usepackage{amssymb}	
\usepackage{booktabs}
\usepackage{array}
\usepackage{ulem}
\usepackage{xcolor}

\bibpunct{(}{)}{;}{a}{}{,}

\title[Sakurai's central star]
{The emergence of a [WC] star in Sakurai's object}

\author[Marcolino et al.]{
W. Marcolino$^{1}$\thanks{E-mail: wagner@astro.ufrj.br},
P. A. M. van Hoof$^{2}$, 
V. C. Maria$^{1}$, 
H. Todt$^{3}$,
G. Van de Steene$^2$,
J. A. Toalá$^4$,
\and \,\,S. Kimeswenger$^5$,
M. Hajduk$^6$,
J.-C. Bouret$^{7}$,
D. Tafoya$^{8}$,
D. Barría$^{9}$
and A. A. Zijlstra$^{10}$\thanks{Corresponding author. Email: albert.zijlstra@manchester.ac.uk}
\\
$^{1}$Observat\'orio do Valongo, Universidade Federal do Rio de Janeiro, Ladeira Pedro Ant\^onio, 43, CEP 20080-090, Rio de Janeiro, Brazil\\
$^{2}$Department of Astronomy \& Astrophysics, Royal Observatory of Belgium, Ringlaan 3, B-1180, Brussels, Belgium\\
$^{3}$ Institut für Physik und Astronomie, Universität Potsdam, Karl-Liebknecht-Str. 24/25, 14476 Potsdam, Germany  \\
$^{4}$ Instituto de Radioastronomía y Astrofísica, Universidad Nacional Autónoma de México, Morelia 58089, México \\
$^{5}$ Institut für Astro- und Teilchenphysik, Leopold–Franzens Universität Innsbruck, Technikerstr. 25, 6020 Innsbruck, Austria \\
$^{6}$ Department of Geodesy, Faculty of Geoengineering, University of Warmia and Mazury, ul. Oczapowskiego 2, 10-719 Olsztyn,
Poland \\
$^{7}$ Aix-Marseille Univ., CNRS, CNES, LAM, 13388 Marseille, France \\
$^{8}$ Department of Space, Earth and Environment, Chalmers University of Technology, Onsala Space Observatory, 439 92 Onsala, Sweden \\
$^{9}$ Centro de Investigación en Ciencias del Espacio y Física Teórica, Universidad Central de Chile, Av. Francisco de Aguirre 0405, La Serena, Chile \\
$^{10}$ Jodrell Bank Centre for Astrophysics, Department of Physics and Astronomy, The University of Manchester, Oxford Road, Manchester M13 9PL, UK \\} 
\date{Accepted XXX. Received YYY; in original form ZZZ}

\pubyear{2026}

\begin{document}
\label{firstpage}
\pagerange{\pageref{firstpage}--\pageref{lastpage}}
\maketitle
\begin{abstract}
{Sakurai’s object provides a rare opportunity to observe stellar evolution on human timescales. Since its born-again event and detection in 1996, its evolution has been extensively monitored, and recent optical spectroscopy has suggested the emergence of [WR]-type emission features. In this Letter, we present a secure spectroscopic identification of its central star based on new VLT/FORS2 observations and NLTE expanding atmosphere models. Several observed emission lines arise in a [WR]-type stellar wind, establishing the central star as a [WR] object. The strongest stellar features are due to \ion{C}{ii-iii} and \ion{He}{i}. We derive the stellar and wind parameters of Sakurai's object and discuss its evolutionary status. The [WR] star is much cooler ($T_\mathrm{eff} \sim$ 30.5 kK) than the born-again star V605 Aql ($T_\mathrm{eff} \sim$ 95 kK), which erupted about 80 years earlier. The inferred temperature is lower than predicted by models suppressing convective mixing efficiency during the VLTP, but is consistent with higher time-resolution calculations involving slightly lower-mass ($\lesssim 0.6$ M$_\odot$) remnants. Continued spectroscopic monitoring and atmospheric modelling will be crucial for tracing its reheating curve and constraining the physics of this rapid evolutionary phase.}
\end{abstract}



\section{Introduction}
\label{sec:intro}

Born-again stars are thought to result from a late helium-shell flash occurring at, or shortly after, the end of the asymptotic giant branch (AGB) phase. Depending on when the pulse occurs, it is classified as an AGB final thermal pulse (AFTP), a late thermal pulse (LTP; post-AGB), or a very late thermal pulse  \citep[VLTP; see][]{bertolami2024}. In the most extreme case (VLTP), the star is already on the white-dwarf cooling track when the flash causes rapid expansion and cooling, accompanied by substantial mass ejection into the circumstellar environment. The star is thus "born again" and returns close to its former AGB position in the HR diagram. Subsequent evolution involves reheating, and possibly an additional cooling excursion, as the star evolves back toward the white-dwarf domain with a markedly altered surface composition \citep*[][]{lawlor2003}.

Only two objects have been directly observed undergoing a VLTP: V605 Aql and Sakurai’s object (V4334 Sgr). Two others are thought to have undergone an LTP: FG Sge \citep[][]{Paczynski1971,vangenderen1995,jeffery2006} and SAO 244567 \citep[][]{reindl2017,lawlor2021-stingray}. FG Sge has evolved to become a cool supergiant. SAO 244567 showed a sudden ionization event around 1980 and is currently cooling. In contrast, V605 Aql, discovered in 1919, evolved from an RCB-like star with $T_\mathrm{eff} \sim 5000$ K in 1921 to a $\sim95,000$ K [WR] star by 2001 \citep[][]{clayton2006}.

Sakurai's object (hereafter SO) has been extensively studied since its discovery in 1996 \citep[][]{nakano-sakurai1996}. Following its outburst, the central star became completely obscured by circumstellar material \citep[see][for a review]{vanhoof2018}. Later, radio and infrared observations continued to reveal its rapid evolution \citep[e.g.,][]{hajduk2005,vanHoof-2007,evans2022sak}. Although some observations were interpreted as evidence for reheating of the central star, the radio emission origin proved to be more complex than initially thought, and no [WR] features had been identified by 2020 \citep[][]{hinkle2020,hajduk2024}. Recently, \cite{griet2025} reported the possible emergence and subsequent strengthening of [WR]-type emission features and proposed tentative identifications for the associated ions. Quantitatively confirming the emergence of the central star and tracking the evolution of its physical parameters can provide invaluable constraints on stellar evolution models. 

In this Letter, we present the first spectroscopic characterization of the central star of SO based on recent VLT/FORS2 observations and NLTE expanding atmosphere models.


\section{Observations and models}
\label{sec:obs}

VLT/FORS2 data were obtained through several campaigns by van Hoof \citep[see][]{griet2025}. Here, we use a 2023 data set (ESO programme 111.24P1.001), which revealed the clearest stellar-like features despite strong nebular emission. The observations were acquired with the 300V+GG435 and 300I+OG590 setups, covering 4245--8870\AA\ and 5980--10310\AA, respectively, with $R\sim350$. Each setup consisted of nine exposures of 920\,s with a 1.3" slit. Details about data reduction can be seen in \cite{vanHoof-2007}. A comprehensive analysis of all available observations will be presented elsewhere.

\begin{figure*}
    \includegraphics[width=\textwidth]{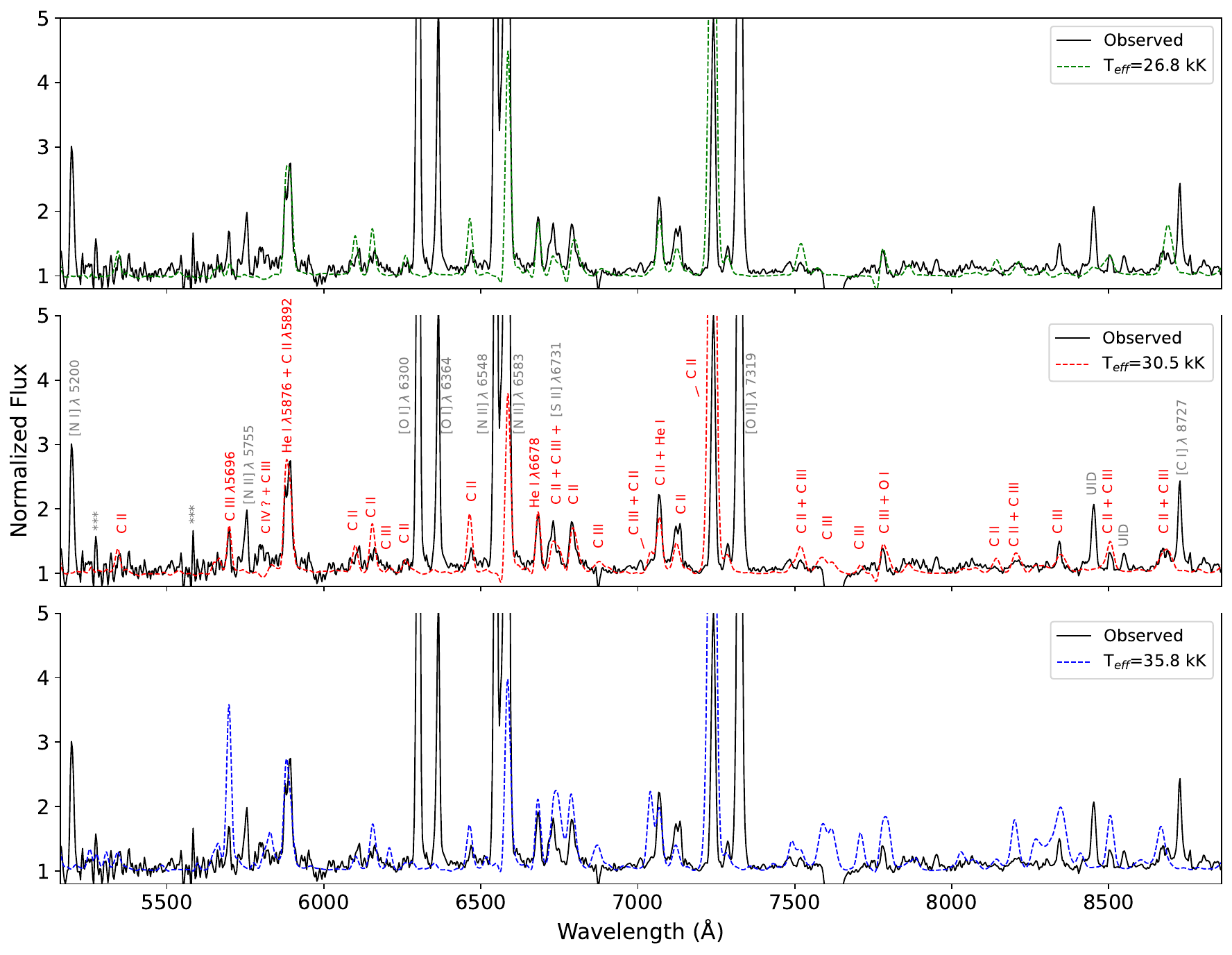}
    \caption{VLT/FORS2 spectrum of Sakurai's object (black solid line; year 2023) and [WR] synthetic spectra with different  temperatures. Known nebular lines, unidentified features (UID), and possible artefacts (asterisks) are indicated (gray). Our fits rule out temperatures above $\sim$ 36 kK and below $\sim$ 27 kK. Note that \ion{C}{iii} lines get stronger when the temperature is increased. Our best model is depicted in the middle panel.} 
    \label{fig:specfit}
\end{figure*}

We used the radiative transfer code CMFGEN \citep*[][]{Hillier98}, which solves the transfer equation in the co-moving frame of an expanding atmosphere coupled to the radiative and statistical equilibrium equations (NLTE). A full spectrum (continuum and lines) can be computed in the observer's frame over a wide wavelength range (far-UV to the mid-IR) and is well suited to modelling WR stars with dense stellar winds. 

The atomic data include hundreds of levels and thousands of transitions from \ion{He}{i-ii}, \ion{C}{ii-iv}, \ion{O}{i-iv}, \ion{Ne}{ii-iv}, \ion{Mg}{ii}, \ion{Al}{ii-iii}, \ion{Si}{ii-iv}, \ion{P}{V}, \ion{S}{iii-iv}, and \ion{Fe}{ii-v}. Test models showed that including or excluding trace ions has little effect on the emergent spectrum. We therefore adopted this set, which provides adequate metal line blanketing without excessive computational cost.

\subsection{Methodology}
\label{sec:meth}

In principle, flux-calibrated data covering a broad wavelength range, combined with a reliable distance estimate, are sufficient to constrain the luminosity through SED fitting. For SO, however, this approach is severely limited.  Distance estimates in the literature span 1.8-8 kpc, but recent studies favour $\sim$2 to 4.4 kpc based, for example, on reddening or  expansion velocity arguments, with 3.5 kpc being the most used value \citep[see the review by][]{hinkle2020}. Nevertheless, no direct distance determination is available. Such an uncertainty, together with the poorly constrained circumstellar extinction and corresponding colour excess \citep[][]{evans2022sak}, introduces strong degeneracies among luminosity, distance, and reddening. Moreover, the observed optical--near-IR flux cannot be straightforwardly attributed to the [WR] central star, as it is likely affected by circumstellar continuum emission and contaminated by strong nebular lines (see below). We therefore restricted our analysis to normalized optical VLT/FORS2 spectra.

During the modelling process, we adopted a luminosity of 5000 $L_\odot$ and an initial temperature guess, from which the stellar radius follows via the Stefan--Boltzmann relation. The mass was fixed at $\sim$0.60 $M_\odot$, as [WR] stars do not exhibit spectroscopic $\log g$ diagnostics. This value is  the observed mean for white-dwarfs \citep[see e.g.,][]{OBrien-WD-masses-GAIA-2024}. We adopted a chemical composition of [He/C/O] $\approx$ [43,51,6] (by mass), typical of [WR] stars and providing a good fit to the observed spectrum (see below). This differs markedly from what is inferred for SO shortly after its outburst, [He/C/O] = [90,7,3] \citep[][]{asplund-sak1999}, and likely results from proton ingestion and burning and mass-loss since then. 

SO is likely associated with the Galactic thick disk \citep{vanHoof-2007}, whose stars typically have sub-solar metallicities \citep[e.g.,][]{bensby2014}. To assess the impact of this, we computed models with reduced metal content, scaling Fe to $\sim$1/4 solar ([Fe/H] $\sim$ -0.6). The resulting spectrum changes only slightly, indicating that our results are not strongly sensitive to plausible variations in the initial metallicity.

The main wind parameters are the mass-loss rate $\dot{M}$ and terminal velocity $V_\infty$, which were adjusted to fit the observations. Clumping was included through a filling-factor formalism ($D=10$), starting at a velocity of $\sim$20\% of $V_\infty$.
In [WR] stars, the stellar temperature primarily determines the wind ionization and thus the dominant ions and relative line intensities. The transformed radius, $R_T \propto R_\star (V_\infty/\dot{M})^{2/3}$, controls the emission-line strengths, while $V_\infty$ sets the line widths. Models with the same temperature and $R_T$ produce similar rectified spectra \citep*[][]{schmutz1989}. We used these properties to iteratively improve the fits and derive a first estimate of the stellar and wind parameters of SO after the born-again event.  A different luminosity can still result in the same best fit, provided that $R_T$ and temperature are kept constant. The radius and wind
parameters would have to change.

\subsection{Results}
\label{sec:results}

Figure \ref{fig:specfit} compares synthetic spectra with the VLT/FORS2 spectrum of SO. Our best model is shown in the middle panel, while the other panels illustrate the effect of varying temperature at fixed $R_T$ (see below). Several \ion{C}{ii-iii} and \ion{He}{i} features are identified.  Those with labelled wavelengths are discussed below. Their identifications were quantitatively confirmed by computing spectra for individual ions\footnote{Initially, we extended the spectral calculations to 9500\AA, where \cite{griet2025} reported the emergence of features possibly associated with carbon (their Fig. 1). We did not find synthetic counterparts to these transitions. They remain unidentified and may instead arise from CN.}. A nebular contribution to the \ion{He}{i} lines is possible but difficult to distinguish at the current spectral resolution. Forbidden nebular lines are also marked in Figure \ref{fig:specfit}. They contaminate several spectral regions, a situation similar to that encountered in the analysis of V605 Aql by \cite{clayton2006}.

The \ion{C}{ii-iii} features provide an estimate of the stellar temperature. The \ion{C}{iii} (\ion{C}{ii}) lines become stronger (weaker) than observed for values above $\sim 30500$ K, while the opposite occurs at lower temperatures. The \ion{C}{iii} $\lambda 5696$ line is particularly sensitive to this effect\footnote{\ion{C}{iii} $\lambda 5696$ is commonly used in spectral classification and gets stronger with temperature from late- to early-type [WR]. The observed FWHM is $\sim$ 10\AA\, and it is compatible with a late-type [WR] \citep[][]{crowther1998-class}.}, as is the \ion{C}{ii-iii} blend located just blueward of the nebular [\ion{C}{i}] $\lambda 8727$ line. Increasing the temperature improves the contribution of \ion{C}{iii} to this blend, shifting the synthetic profile toward the blue side of the observed feature, while decreasing the temperature strengthens the \ion{C}{ii} component and shifts the emission toward the red side of the profile. The best agreement is obtained near $T_\mathrm{eff} \sim 30500$ K. 

The feature around $\lambda 5800$, redward of [\ion{N}{ii}] $\lambda 5755$, may be associated with \ion{C}{iv}. In our models, however, this feature appears only at $T_\mathrm{eff}$'s well above $\sim 30500$ K, where the neighboring \ion{C}{iii} emission becomes too strong compared to the observations. This behaviour is illustrated by the $T_\mathrm{eff} \sim 36000$ K model in the bottom panel of Fig. \ref{fig:specfit}. Models with $T_\mathrm{eff}$'s up to $\sim 42000$ K produce conspicuous \ion{C}{iv} emission, but \ion{C}{iii} lines become excessively strong and the \ion{C}{ii} features too weak, resulting in an unacceptable fit.

Some discrepancies in the line intensities, as well as the difficulty in reproducing the feature near $\lambda 5800$, could not be solved despite our efforts. For example, we have included X-rays, which are expected from shocks arising from instabilities within a stellar wind. However, it had virtually no impact on the optical spectrum. 

The inability to simultaneously reproduce all observed features may indicate that the atmospheric ionization structure (e.g., the \ion{C}{ii-iii-iv} balance) is not fully captured by our models. This could arise, for example, if the emerging wind of SO is not symmetric, as assumed here. Surface rotation, for example, could break the spherical symmetry and produce latitudinal-dependent wind properties. Furthermore, the central star is not detected in direct imaging, and its optical spectrum may be observed through scattering by the surrounding dusty material, as suggested for V605 Aql \citep[][]{clayton2006}. We cannot rule out that scattering modifies the observed [WR] spectrum. The weak feature near $\lambda 5800$ may also have an identification other than \ion{C}{iv}. Future observations will be essential to clarify these issues, particularly as the central star continues to reheat and \ion{C}{iv} transitions may become prominent.

Despite the low spectral resolution of the VLT/FORS2 data,  nebular contamination, and the blending of several features, we can state conservatively that the current stellar temperature of SO lies in the range $27000\,\mathrm{K} \lesssim T_\mathrm{eff} \lesssim 36000\,\mathrm{K}$. The remaining parameters of our preferred model are summarized in Table \ref{table:stellarparameters}.

\begin{table}
\centering
\caption{[WR] model parameters for Sakurai's object.}
\label{table:stellarparameters}

\begin{tabular}{lrl}
\hline
Parameter & Value & Comments \\
\hline
$\log (L/L_\odot)$ & 3.7 & Adopted \\
Clumping factor, $D$   & 10  & Adopted \\
$R_\ast$ ($\tau =2/3$) & 2.5 $R_\odot$ & $R_\ast \propto L^{1/2}$ \\
$T_\mathrm{eff}$ ($\tau =2/3$) & 30\,550 K & \\
$v_\infty$ (km~s$^{-1}$) & 500 & \\
$\dot{M}$ ($M_\odot\,{\rm yr}^{-1}$) & $1.1 \times10^{-6}$ & $\dot{M}
\propto D^{-1/2}L^{3/4}$ \\
He, C, O, other (solar) & 0.425, 0.510, 0.059, 0.005 & mass fractions\\
\hline
\end{tabular}
\end{table}

\section{Evolutionary status}
\label{sec:itsevolutionbaby}

\begin{figure}
	\includegraphics[width=\columnwidth]{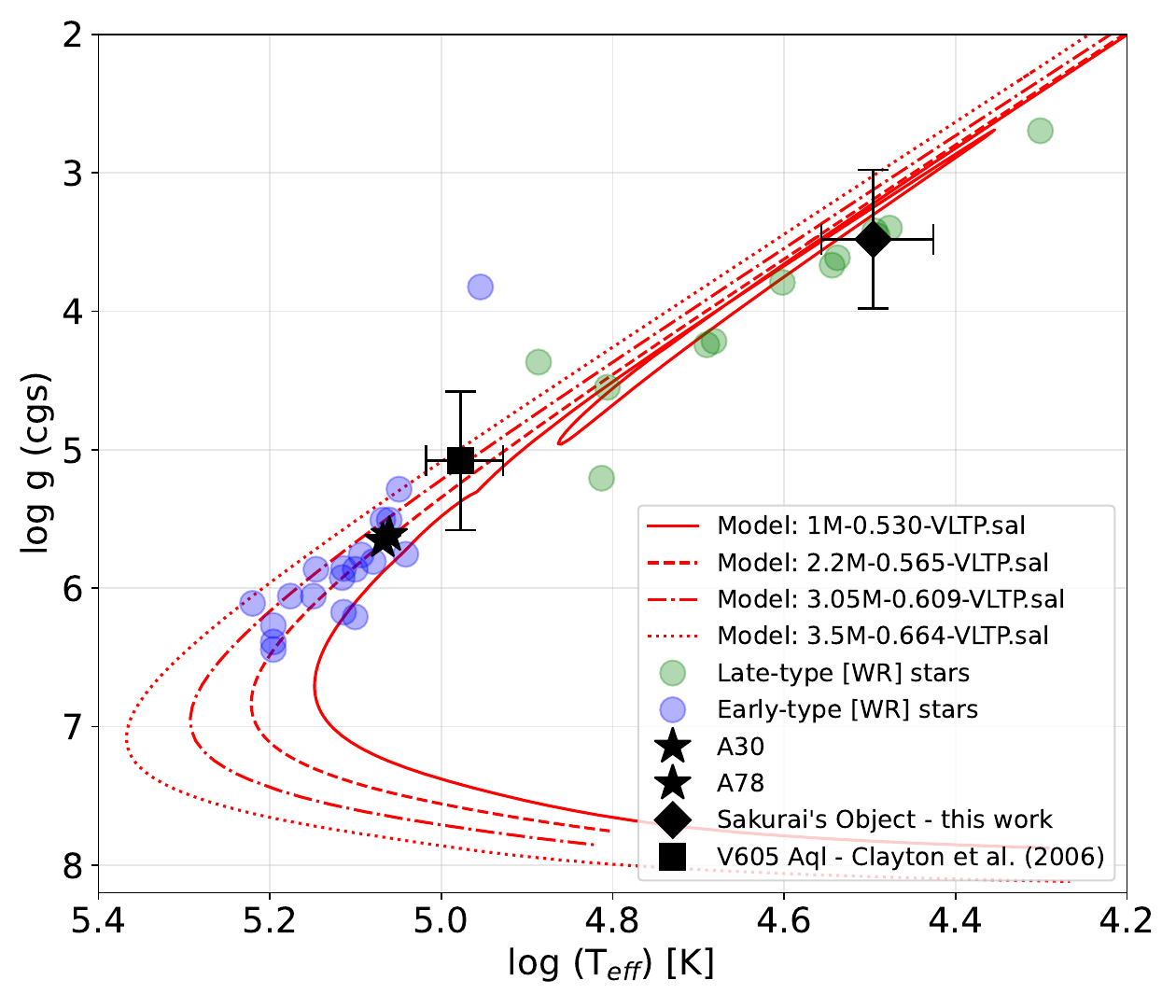}
    \caption{Kiel diagram with evolutionary tracks from \citet{bertolamitracks2006}. 
    SO position is indicated by a black diamond. Late and early-type [WR] stars are also shown. The [WCE] outlier is PN RaMul 2.} 
    \label{fig:kiel}
\end{figure}

Having inferred the stellar and wind parameters of SO, we can now discuss its evolutionary status. Figure \ref{fig:kiel} shows a Kiel diagram including post-VLTP evolutionary tracks from \cite*{bertolamitracks2006} for initial (remnant) masses of 1 (0.53), 2.2 (0.565), 3.05 (0.609), and 3.5 M$_\odot$ (0.664 M$_\odot$). The position of SO is indicated by a black diamond, while V605 Aql is shown as a black square based on the results of \cite{clayton2006}. We also include the more evolved born-again stars A 30 and A 78 \citep[][]{guerrero2012,toala2015}. For SO, the temperature uncertainty reflects the conservative range inferred in Section \ref{sec:results}. The surface gravity was computed assuming $M=0.6\,M_\odot$ (see Section \ref{sec:meth}). Adopting masses between $0.5$ and $0.7\,M_\odot$ produces only minor shifts in the diagram.  We estimated an uncertainty of $\pm0.5$ dex in $\log g$ by propagating the uncertainties in the derived $T_\mathrm{eff}$ and in the adopted luminosity. We assumed a 50\% uncertainty in the distance for the latter.

We also added results from the literature for early- and late-type [WR] stars in Figure \ref{fig:kiel} using data from \cite{koesterke2001}, \cite{marcolino07}, \cite*{keller2014}, \cite{gomez-gonzalez2022}, and \cite*{toala2024}. When the same object was analyzed in more than one study, we kept the latest result. V605 Aql is closer to early-type objects, whereas SO position is among late-type [WR] stars, suggesting a younger evolutionary status. 

\begin{figure}
	\includegraphics[width=\columnwidth]{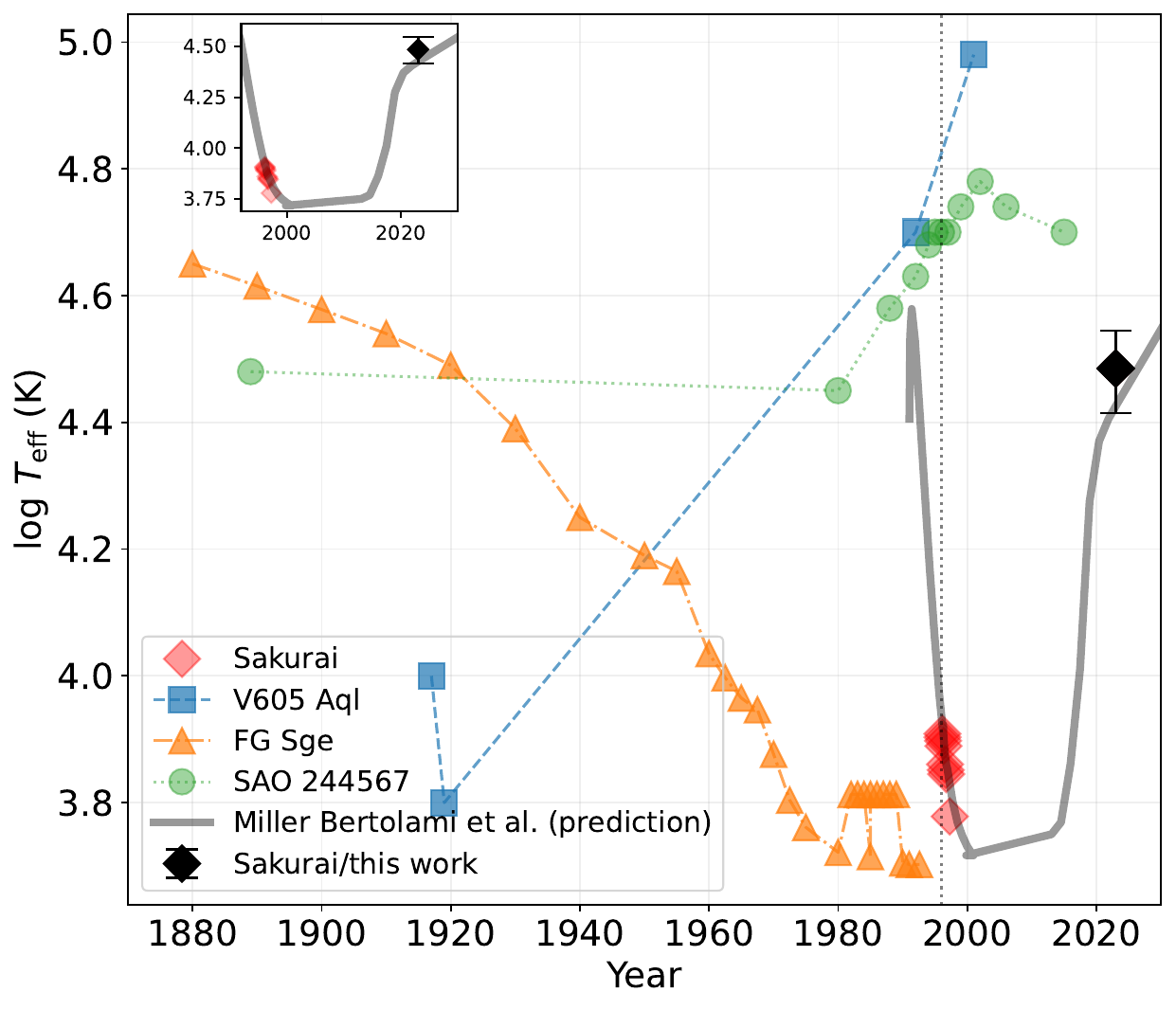}
    \caption{Evolution of the temperature of the stars undergoing the born-again phenomenon: SO, V605 Aql, SAO 244567, and FG Sge.}
    \label{fig:realtime}
\end{figure}

 Figure \ref{fig:realtime} shows the temperature evolution of SO, V605 Aql, SAO 244567, and FG Sge, using data from \cite{vangenderen1995}, \cite{lawlor2003}, \cite{schaefer2015}, and \cite{reindl2014, reindl2017}. For SO and V605 Aql, the most recent points correspond to our results and those of \cite{clayton2006}, respectively.  We also display the model prediction by \cite{bertolami2006} for a 0.5842 M$_\odot$ remnant star (see below). Our SO data point is the first after 1998.

The rapid cooling of SO can be reasonably reproduced by theoretical calculations in the literature, as well as its brightening time after the VLTP up to its discovery in 1996, the born-again time, which is about 5 yr \citep[see e.g.][]{hajduk2005, bertolami2006}. However, the calculations of \cite{herwig2001} and \cite{hajduk2005} required a substantial reduction in the convective mixing efficiency within the He-flash zone; otherwise, the born-again timescale would be about 350 yr \citep*[see][]{herwig2001}. In contrast, \cite{bertolami2006} and \cite*{bertolami2007} showed that high temporal resolution ($\sim$$10^{-5}$ yr) and slightly lower remnant masses ($\sim$0.58 M$_\odot$) were sufficient to reproduce the rapid evolution of SO without artificially reducing the mixing efficiency.

Regarding the reheating of SO, \cite{hajduk2005} predicted $\log T_\mathrm{eff}\sim 4.95$ ($\sim$ 90 kK) by 2020, implying a much faster temperature increase than supported by our results. On the other hand, the model of \cite{bertolami2006} is broadly consistent with our measurements (see Fig. \ref{fig:realtime}). In their calculations, temperatures above $\log T_\mathrm{eff}\sim 4.80$ ($\gtrsim$ 60 kK) are reached only about 100 years after the VLTP, while $\log T_\mathrm{eff}\sim 4.50$ is attained after $\sim$ 35 years (see also their Fig. 5). However, these authors noted that the timescales involved depend critically on the numerical treatment adopted once the models reach the Eddington limit.

\section{Discussion}
\label{sec:discussion}

The analysis presented in Sect. \ref{sec:results} strongly suggests that several of the observed features originate in a stellar wind. Although some discrepancies remain, their relative intensities are reasonably well reproduced by our models. Nevertheless, one could argue that these features are instead formed in the ejecta produced during the VLTP event rather than in a stellar wind. Given the low spectral resolving power of the observations, synthetic wind-line profiles convolved to the instrumental resolution could potentially mimic emission features produced elsewhere, leading to an apparently satisfactory fit by coincidence.

The circumstellar material of SO has been mapped in the infrared with adaptive optics by \cite{hinkle2020} and at radio wavelengths with ALMA by \cite{tafoya2023}. These observations revealed an equatorial disk and bipolar outflows forming an hourglass geometry. The projected velocities in the bipolar flows reach several hundred km s$^{-1}$ \citep[see Fig. 3 of][]{tafoya2023}, comparable to the terminal velocity of our best model ($V_\infty = 500$ km s$^{-1}$). However, it would be a remarkable coincidence if a spherically symmetric NLTE expanding atmosphere model reproduced the relative intensities of several observed features were they formed predominantly in the asymmetric ejecta rather than in a stellar wind.

\begin{figure}
	\includegraphics[width=\columnwidth]{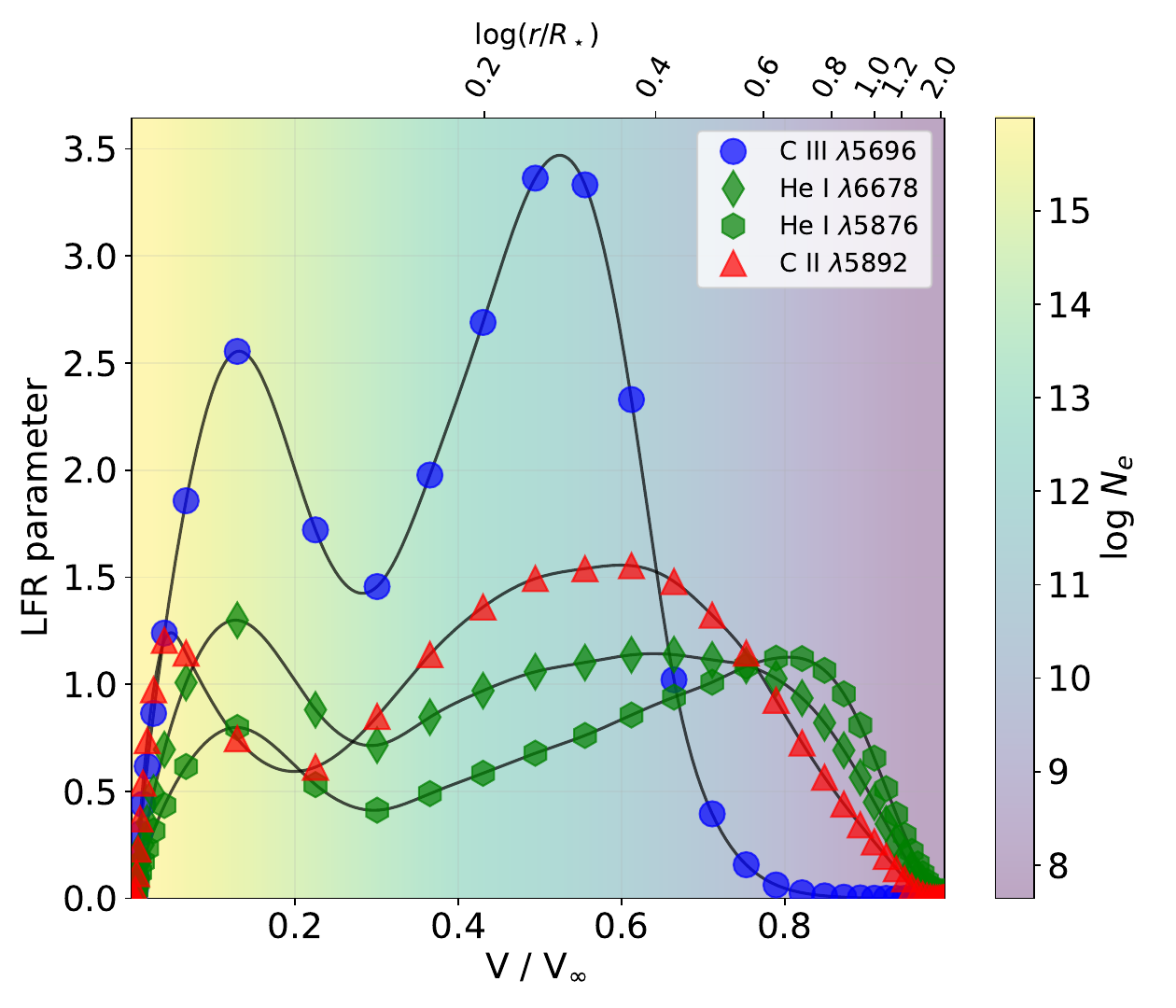}
    \caption{Line formation regions (LFRs) of selected lines. Note that the lines are not formed in a specific point, but over a large velocity and density interval. { The LFR parameter informs how much of the profile is formed at a certain point \citep[see][for more details]{hillier-LFR-1987}.}} 
   \label{fig:lineform}
\end{figure}

The velocity, density, and temperature structure of a [WR] wind are expected to differ significantly from those of the born-again ejecta. Figure \ref{fig:lineform} illustrates the formation regions of selected \ion{C}{ii-iii} and \ion{He}{i} transitions in our best model. The horizontal axis shows the wind velocity normalized to the terminal velocity, while the left vertical axis indicates the contribution to the line profile at a given velocity  \citep[for the quantitative definition of the LFR parameter, see][Sect.\ion{}{iv}]{hillier-LFR-1987}; the right axis shows the electron density. The lines form over most of the wind volume, spanning a wide range of velocities and densities. Regions above $\sim$80\% of $v_\infty$ still contribute to the \ion{He}{i} and \ion{C}{ii} lines, whereas \ion{C}{iii} forms mainly in the inner wind. This extended formation region reflects the stratified structure and accelerating velocity field of a WR wind, conditions that would be difficult to reproduce in the ejecta while yielding comparable line intensities.

It is also instructive to compare the spatial scales implied by the ALMA observations with those of our atmosphere model. From H$^{12}$CN data, the ejecta reach velocities of $\sim$300 km s$^{-1}$ at $\sim$600 AU. In contrast, our model extends to $\sim$ 200 stellar radii, where the velocity reaches $\sim$500 km s$^{-1}$, corresponding to only $\sim$2 AU. The entire [WR] spectrum is formed within this compact region, indicating that the line-forming zone probes a physical regime entirely distinct from the large-scale ejecta.

Note also that the \ion{C}{ii-iii} features identified in our models, for example, are not produced under the same physical conditions as the nebular lines [\ion{N}{i-ii}], [\ion{O}{i-ii}], and [\ion{C}{i}]. These forbidden transitions trace neutral or weakly ionized, low-density gas, whereas the \ion{C}{ii-iii} recombination features require a denser and hotter environment. In particular, the upper level of \ion{C}{iii} $\lambda 5696$ can be efficiently populated only in dense stellar winds \citep[][]{hillier1989}. 

In summary, our analysis favors a stellar-wind origin for the broad \ion{C}{ii-iii} features. The inferred stellar parameters correspond to a late-type [WR] object (see Section \ref{sec:itsevolutionbaby}). The forbidden lines trace the surrounding ejecta.

\section{Conclusions}
\label{sec:conclusions}

Our VLT/FORS2 observations and NLTE models confirm the emergence of a [WC]-type spectrum in SO, first reported by \cite{griet2025}, approximately 35 yr after its VLTP event. The relative intensities of several optical emission lines are reasonably reproduced by our synthetic spectra, indicating that most originate from \ion{C}{ii-iii} and \ion{He}{i}. The \ion{C}{ii-iii} lines constrain the current  temperature to a conservative range of 27000 K $\lesssim T_{\rm eff} \lesssim 36000$ K. The parameters of our preferred model are summarized in Table \ref{table:stellarparameters}. Our results support a [WCL] classification for SO, placing it at an earlier evolutionary stage than V605 Aql. The inferred temperature is significantly lower than predicted by models that suppress the convective mixing efficiency \citep[][]{hajduk2005}, but broadly consistent with the calculations of \cite{bertolami2006}.

Continued spectroscopic monitoring of SO remains essential. Born-again objects provide a rare opportunity to observe stellar evolution in real time and place unique constraints on VLTP evolution and the emergence of hydrogen-deficient central stars.

\section*{Acknowledgements}

Based on observations collected at the European Southern Observatory under ESO programme 111.24P1.001.

\section*{DATA AVAILABILITY}
The data underlying this article are available from the cited sources or by e-mail request to wagner@ov.ufrj.br.

\bibliographystyle{mnras}
\bibliography{biblio}







\bsp	
\label{lastpage}
\end{document}